\def\changesHilighted{false} 

\documentclass[conference,a4paper]{APSIPA2023}
\usepackage{multirow}
\usepackage{graphicx}
\usepackage{amsmath}
\usepackage{amssymb}
\usepackage{amsxtra}
\usepackage{threeparttable}
\usepackage{balance}
\usepackage{siunitx}

\usepackage{geometry}
\usepackage{fancyhdr}

\fancypagestyle{firststyle} {
    \fancyhf{}
    \fancyhead[C]{2023 Asia Pacific Signal and Information Processing Association Annual Summit and Conference (APSIPA ASC)}
}

\usepackage{amsthm}
\newtheorem{problem}{Problem}
\newtheorem{remark}{Remark}

\usepackage{mathtools}

\usepackage{dblfloatfix}    

\usepackage{cite}

\mathtoolsset{showonlyrefs=true}
\usepackage{color}
\definecolor{forestgreen}{rgb}{0.33,0.61,0.34}
\newcommand{\add}[1]{\textcolor{blue}{#1}}
\newcommand{\masaki}[1]{\textcolor{forestgreen}{\bf [Masaki: #1]}}
\newcommand{\aizawa}[1]{\textcolor{blue}{\bf [Aizawa: #1]}}
\newcommand{\waka}[1]{\textcolor{red}{\bf [Waka: #1]}}
\newcommand{\edi}[1]{\textcolor{brown}{\bf [Edi: #1]}}

\ifnum\pdfstrcmp{\changesHilighted}{false}=0
    \renewcommand{\aizawa}[1]{}
    \renewcommand{\masaki}[1]{}
    \renewcommand{\waka}[1]{}
    \renewcommand{\edi}[1]{}
    \renewcommand{\add}[1]{#1}
\fi

\usepackage{algorithmic}
\usepackage{algorithm}

\usepackage{diagbox}
\usepackage{subcaption}

\begin{document}

\title{Manipulation of Neuronal Network Firing Patterns using Temporal Deep Unfolding-based MPC}

\author{%
\authorblockN{%
Jumpei Aizawa\authorrefmark{1}, 
Masaki Ogura\authorrefmark{1}, 
Masanori Shimono\authorrefmark{2} and
Naoki Wakamiya\authorrefmark{1}
}
\authorblockA{%
\authorrefmark{1}
Osaka University, Japan  \\
E-mail: \{j-aizawa, m-ogura, wakamiya\}@ist.osaka-u.ac.jp  Tel: +81-06-6879-4358}
\authorblockA{%
\authorrefmark{2}
Kyoto University, Japan \\
E-mail: shimono.masanori.7w@kyoto-u.ac.jp  Tel: +81-075-751-4173}
}

\maketitle
\thispagestyle{firststyle}
\pagestyle{fancy}

\begin{abstract}
Because neuronal networks are intricate systems composed of interconnected neurons, their control poses challenges owing to their nonlinearity and complexity. In this paper, we propose a method to design control input to a neuronal network to manipulate the firing patterns of modules within the network. We propose a methodology for designing a control input based on temporal deep unfolding-based model predictive control (TDU-MPC), a control methodology based on the deep unfolding technique actively investigated in the context of wireless signal processing. During the method development, we address the unique characteristics of neuron dynamics, such as zero gradients in firing times, by approximating input currents using a sigmoid function. The effectiveness of the proposed method is confirmed via numerical simulations. 
{In networks with 15 and 30 neurons, the control was achieved to switch the firing frequencies of two modules without directly applying control inputs.}
This study includes a tailored methodology for networked neurons, the extension of TDU-MPC to nonlinear systems with reset dynamics, and the achievement of desired firing patterns in neuronal networks.
\end{abstract}

\newcommand\blfootnote[1]{%
  \begingroup
  \renewcommand\thefootnote{}\footnote{#1}%
  \addtocounter{footnote}{-1}%
  \endgroup
}
\blfootnote{This work was supported by JSPS KAKENHI Grant Numbers JP22H00514 and JP21H01352.}

\section{Introduction}

The brain, a highly complex organ, consists of a vast network of interconnected neurons. When an individual neuron within the network generates an electrical impulse called firing, neighboring neurons receive an electrical influx that serves as a stimulus to trigger their firing. This phenomenon propagates throughout the entire network and contributes to the complex pattern of the firing dynamics in the brain. Hence, the brain exhibits a wide range of electrically active patterns and dynamic behaviors that account for its remarkable capabilities and functions~\cite{hutchison2013dynamic}.

Studies focused on brain dynamics control~\cite{tang2018colloquium} have the potential to yield significant advancements in the enhancement of human cognitive function and the performance of brain--machine interfaces and artificial nerves. For example, advancements in brain dynamics control are expected to pave the way for the development of novel therapies for psychiatric disorders~\cite{lee2019current}. In addition, a greater degree of seamless communication between the human brain and external devices can be achieved by precisely controlling the firing patterns of neurons to optimize the brain--machine interfaces~\cite{wander2014brain}. The prospect of using brain dynamics control to improve human welfare, both in terms of cognitive enhancements and psychiatric disorder treatments, demonstrates the potential of such research~\cite{tang2018colloquium}.

Owing to the complex nature of a neuronal network, controlling its firing pattern is challenging ~\cite{iolov2014stochastic}. A neuron is often modeled as a nonlinear dynamical system characterized by abrupt jumps in membrane potential as neurons are fired~\cite{izhikevich2003simple}. Such nonlinearity adds a layer of complexity to the control problem~\cite{motter2015networkcontrology}. In addition, a neuronal network as a whole represents a large system consisting of several interconnected neurons with complex interactions. Network size further complicates control efforts by introducing a large number of variables and inter-dependencies, which are to be considered. Addressing these challenges requires developing sophisticated control strategies capable of accommodating the unique characteristics and complexities inherent to neuronal network dynamics.

In this study, we aim to develop a methodology for designing control input currents to effectively manipulate the firing frequency of multiple modules within a network, even when neurons not belonging to the modules are under control. 
Therefore, we employed temporal deep unfolding-based model predictive control (TDU-MPC) introduced in~\cite{kishida2021} as a control methodology. TDU-MPC offers solutions for tackling optimization problems that arise during each step of MPC by employing emerging technologies, such as deep unfolding~\cite{hershey2014deep,Wadayama,Monga,Melodia,Ogura}. Deep unfolding involves the temporal unfolding of the mathematical model of a control object by treating it as a feed-forward network. Using backpropagation, the control inputs at each instant of time step are acquired via learning, and the iterative process is repeated for each step. 

Although TDU-MPC has been recognized for its effectiveness in the control of nonlinear dynamical systems~\cite{kishida2021}, the dynamic characteristics of neurons pose challenges. The main challenge lies in the fact that once a neuron model is discretized, the firing times of the neurons \add{can} have a zero gradient with respect to the control\add{, as has been observed by our preliminary experiments}.\edi{参考文献を記載}\masaki{done}
\add{This property of the gradient} complicates the process of learning control inputs because TDU-MPC relies on deep learning techniques. To overcome such hurdles, we developed an approach to approximate the input current of neighboring neurons by means of a sigmoid function, which enabled us to avoid the aforementioned problem of zero gradient. 

The contributions of this study are summarized as follows. First, the paper proposes the methodology for designing control input to neuronal networks to achieve a desired firing pattern of modules within the networks. Second, unlike other methodologies primarily focused on single neurons~\cite{iolov2014stochastic}, our methodology is tailored to \emph{networks} of neurons. Third, the applicability of TDU-MPC can be extended to include nonlinear systems exhibiting reset dynamics. 

This paper is organized as follows: Section~\ref{sec:problem} states the problem studied in this paper, while Section~\ref{sec:proposed} proposes a method for solving the control problem. Section~\ref{sec:result} presents the results of numerical simulations. Finally, the paper is concluded in Section~\ref{sec:conclusion}.

\section{Problem Statement}\label{sec:problem}

In this section, we formally state our problem of controlling the neuronal dynamics. In Subsection~\ref{subsec:net}, we state our model of neuronal networks. Subsequently, in Subsection~\ref{control_obj}, we state the control problem addressed in this paper.  

\subsection{Neuronal Network}\label{subsec:net}

In this subsection, we state the model of a neuronal network. Let $\mathcal G = (\mathcal V, \mathcal E)$ denote the directed graph representing the connectivity of the neuronal network. The node set~$\mathcal V = \{1, \dotsc, N\}$ represents the set of neurons, while the edge set~$\mathcal E$ represents the connectivity among the neurons. For each $i \in V$, $\mathcal N_i$ denotes the set of in-neighbors of node~$i$. Under this notation, we assume that when a neuron in~$\mathcal V_i$ fires, the neuron~$i$ receives a current. 

\newcommand{\exitatory}{\textrm{ex}}
\newcommand{\inhibitory}{\textrm{in}}
\newcommand{\ctr}{\textrm{control}}
\newcommand{\internal}{\textrm{internal}}

\add{The type of firing neuron determines the effect on its neighboring neurons.}\masaki{この追加の意図はなに？}\aizawa{元の文とeditageで意味が変わっていたので，よりわかりやすい（と思われる）文に直しました．}\masaki{OK}
For this study, we assume, as is common in the literature, that neurons are divided into two types of neurons: excitatory neurons and inhibitory neurons. The set of the former neurons is denoted by~$\mathcal V_{\exitatory}$, while the latter is denoted by~$\mathcal V_{\inhibitory}$. An excitatory neuron induces a positive current~$I_{\exitatory}$ to its neighbors, while an inhibitory neuron induces a negative current~$I_{\inhibitory}$. 

We assume that each neuron $i\in \mathcal V$ follows the Izhikevich model~\cite{izhikevich2003simple} described by the differential equation
\begin{align}
\dot v_i(t) &= 0.04v_i^2(t) + 5v_i(t) + 140 - u_i(t) + I_i(t) \label{Izhi:1}, \\
\dot u_i(t)&= a(bv_i(t) - u_i(t)) \label{Izhi:2}
\end{align}
with the reset dynamics described by 
\begin{equation}
\label{Izhi:3}
\begin{bmatrix}
    v_i(t^+)\\u_i(t^+)
\end{bmatrix}
=
\begin{bmatrix}
    c \\ u_i(t)+d
\end{bmatrix},\ 
\mbox{ if } v_i(t) \geq 30.
\end{equation}
For this study, we assume that a neuron~$i$ fires at time~$t$ if $v_i(t) \geq 30$. The real numbers~$a$, $b$, $c$, and~$d$ in the dynamical model are considered dimensionless parameters.
By varying the values of these dimensionless parameters, a rich variety of neuronal behaviors, including regular spiking, bursting, and fast spiking, can be achieved. Therefore, the Izhikevich model is extensively used to study and replicate various types of neuronal activity seen in different regions of the brain. 

In the differential equations~\eqref{Izhi:1} and~\eqref{Izhi:2}, 
the scalar variable~$v_i(t)$ represents the membrane potential of neuron~$i$ at time~$t$, while $u_i(t)$ represents the value of the membrane recovery variable of neuron~$i$ at time~$t$. Further, $I_i(t)$ represents the current input to neuron~$i$ and is assumed as 
\begin{equation}
    I_i(t) = I_{i, \ctr}(t) + I_{i, \internal}(t), 
\end{equation}
where $I_{i, \ctr}(t)$ is the control input signal that we need to design, and~$I_{i, \internal}(t)$ represents the inter-network current induced by firing events of neurons. In this paper, for simplicity, the inter-network current is supposed to be of the additive form defined as 
\begin{equation}\label{eq:Iinternal(t)}
    I_{i, \internal}(t) = \sum_{j\in \mathcal N_i} I_{ij}(t), 
\end{equation}
where $I_{ij}(t)$ represents the current input from neuron~$j$ to neuron~$i$ and is assumed to be of the form
\begin{equation}\label{eq:I_ij(t)}
    I_{ij}(t) = \begin{cases}
        I_{\exitatory},&\mbox{if neuron~$j$ fired in~$[t-\tau, t]$ and~$j\in \mathcal V_{\exitatory}$}, 
        \\
        I_{\inhibitory}, &\mbox{if neuron~$j$ fired in~$[t-\tau, t]$ and~$j\in \mathcal V_{\inhibitory}$},
        \\
        0,&\mbox{otherwise, }
    \end{cases}
\end{equation}
where $\tau > 0$ is a parameter.

\subsection{Control Objective}\label{control_obj}

For this study, we consider the selective and dynamic enhancement problems caused by the firing activity of a specified module of neurons. Specifically, the network is assumed to consist of three modules, with stronger within-module connectivities than inter-module connectivities. Further, we assume that the control current is applied only to the neurons within a specific module. \add{We consider a closed-loop control of the neurons; therefore, we assume that we can use the membrane potentials~$v_i(t)$ for all $i\in \mathcal V$ to determine the input current $I_{i,\ctr}(t)$ for $i\in \mathcal V_{\ctr}$.}
Under this supposition, our objective is to control the firing frequency of other modules.\waka{何が観測可能か、も仮定として説明する必要があるのでは}
Specifically, we aim to design the following firing pattern: 
From the initial to a pre-specified time, a module~$\mathcal V_1 \subset \mathcal V$ fires frequently while another module~$\mathcal V_2 \subset \mathcal V$ exhibits low firing activities. Thereafter, the module~$\mathcal V_2$ should fire frequently, while module~$\mathcal V_1$ fires less frequently. 

Thus, the aforementioned control problem can be formally stated as follows:

\begin{problem}\label{prf:}
Assume that the set~$\mathcal V$ of neurons is divided into disjoint subsets $\mathcal V_{\ctr}$, $\mathcal V_1$, and~$\mathcal V_2$. Assume 
\begin{equation}
I_{i, \ctr}(t) = 0\mbox{  for all $i \in \mathcal V_1 \cup \mathcal V_2$}. 
\end{equation}
Let $T_s$ and~$T_e$ be real numbers satisfying $0<T_s<T_e$ and define intervals~$\mathcal I_1$ and~$\mathcal I_2$ by 
\begin{equation}
\mathcal I_1= [0, T_s],\ \mathcal I_2 = [T_s, T_e]. 
\end{equation}
For a subset~$\mathcal W$ of~$\mathcal V$ and a time interval~$\mathcal I$, define  
\begin{equation}
    f_{\mathcal W, \mathcal I} = \sum_{i \in \mathcal \mathcal W} \mbox{(number of fires of neuron~$i$ on $\mathcal I$)}.
\end{equation}
\add{Construct a feedback controller of the form}
\begin{equation}\label{eq:controller}
    \add{I_{i, \ctr}(t) = K_i(v_1(t),\dotsc, v_N(t)})
\end{equation}
\add{for all $i\in \mathcal V_{\ctr}$}
such that \add{the objective function}
\begin{equation}\label{eq:obj}
    (f_{\mathcal V_1, \mathcal I_1}-f_{\mathcal V_2, \mathcal I_1})
    +
(f_{\mathcal V_2, \mathcal I_2}-f_{\mathcal V_1, \mathcal I_2})
\end{equation}
is maximized.     
\end{problem}


\begin{remark}
\add{In the practical implementation of the control rules~\eqref{eq:controller}, real-time calculation of the control input~$I_{i,\ctr}$ is preferred. However, in this research, real-time capability is not a requirement, and its realization is left for future investigation. Similarly, the assumption of membrane potential observability, though challenging in practice, aligns with existing practice in the literature (see, e.g., \cite{iolov2014stochastic}).}
\end{remark}


\begin{figure*}[tb]
\centering
  \includegraphics[keepaspectratio, width=.77\linewidth]{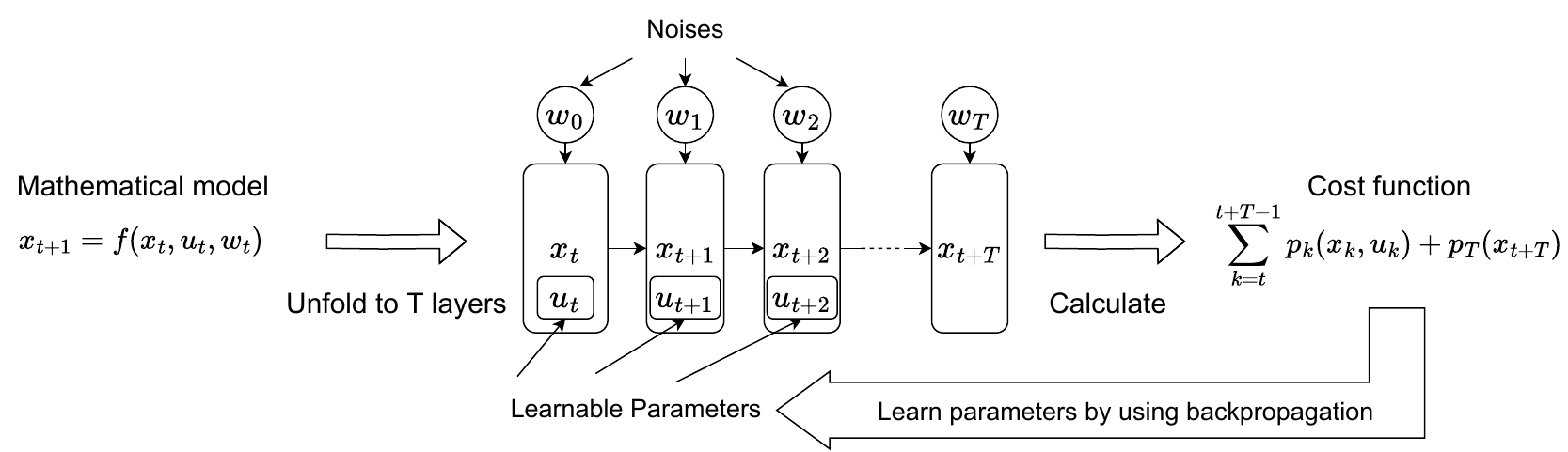}
  \caption{Schematic picture of temporal deep unfolding-based model predictive control (TDU-MPC)}
 \label{temporaldeepunfoldingMPC}
\end{figure*}

Because Problem~\ref{prf:} requires the control of a nonlinear reset system\cite{carrasco2010passivity}, a standard and effective solution involves the use of model predictive control (MPC)\cite{garcia1989model}. MPC is a type of \add{feedback}\masaki{ナイス} control method that aims to achieve optimal control by predicting the behavior of a controlled object at a fixed interval in the future from the current time. In MPC, the controlling input at each instance is determined by solving an optimization problem. The optimization problem consists of a mathematical model of the control target, a cost function obtained from the outputs in the prediction horizon, and constraints. Owing to its effectiveness in optimizing control inputs, handling constraints, and dealing with nonlinear systems, MPC has various applications, including process control, energy management, and automotive systems. Among various approaches available for performing MPC, we chose to use TDU-MPC~\cite{kishida2021} for this study. TDU-MPC offers several advantages such as easy implementation and the ability to find high-quality solutions.

\section{Proposed Method}\label{sec:proposed}

In this section, we propose a method to solve Problem~\ref{prf:} leveraging TDU-MPC. In Subsection~\ref{subsec:TDUMPC}, we present an overview of TDU-MPC, while Subsection~\ref{subsec:discrete-time} presents the proposed methodology for solving Problem~\ref{prf:}.

\subsection{TDU-MPC}\label{subsec:TDUMPC}

To understand TDU-MPC, we begin by reviewing deep unfolding~\cite{hershey2014deep,Wadayama,Monga,Melodia,Ogura} that TDU-MPC is built upon. Deep unfolding is a method for tuning the parameters of iterative algorithms using techniques from deep learning. In deep unfolding, a given iterative algorithm is first unfolded in the time direction and considered a feed-forward network with the number of layers equal to the number of iterations. Next, the parameters of each step of the iterative algorithm are embedded as learnable parameters in the network and tuned via backpropagation and stochastic gradient descent methods. Incremental learning, a widely used technique, is adopted to avoid gradient loss that can occur during parameter learning.

TDU-MPC has been proposed as a method to overcome the disadvantages of existing methodologies for performing MPC. In this subsection, we provide a brief overview of TDU-MPC, considering a discrete-time dynamical system such as 
\begin{equation}
\label{MPCmodel}
    \Sigma : x[k+1]=f(x[k],u[k],w[k])\add{,\ k=0, 1, 2, \dotsc}
\end{equation}
\add{where $x$, $u$, and~$w$ represent the state variable, control input, and disturbance signal, respectively.}\masaki{done}
In TDU-MPC, the system is considered as an iterative algorithm, and the control input is regarded as a learnable parameter.

\edi{文章の意味が不明}\masaki{OK}
\add{Specifically, we first make the feed-forward network with $T$ layers. Each layer is the state $x$ at each instance of time of unfolded $T$-step state transitions in the temporal direction.}
Next, the control inputs at each instance can be learned using deep learning techniques as in deep unfolding.
The disturbances at each instance of time are used as data, and learning is performed using the objective function of the optimization problem in MPC as the cost function.
Thereafter, the control input minimizing the following objective function 
\begin{equation}
\label{cost}
    \sum_{\ell=0}^{T-1}p_\ell(x[k+\ell], u[k+\ell])+p_T(x[k+T])
\end{equation}
is designed using the aforementioned scheme.
In Fig.~\ref{temporaldeepunfoldingMPC}, we illustrate a schematic picture illustrating TDU-MPC. For the details of TDU-MPC, readers are referred to~\cite{kishida2021}.

\subsection{Discrete-time Model}\label{subsec:discrete-time}
As described in the previous subsection, TDU-MPC is a methodology for controlling discrete-time dynamical systems, while Problem~\ref{prf:} involves a continuous-time model. Therefore, for applying TDU-MPC to Problem~\ref{prf:}, the dynamics presented as~\eqref{Izhi:1}--\eqref{Izhi:3} and the objective function~\eqref{eq:obj} must be appropriately discretized. 

Let $\Delta t > 0$ denote the time step for discretization. For this study, we discretized the differential equations~\eqref{Izhi:1} and~\eqref{Izhi:2} as 
\begin{equation}
\label{discrete:1}
        v_i[k+1]
    =    
    \begin{cases}
        v_i[k] + (0.04v_i^2[k] + 5v_i[k] \\
        \quad\quad+ 140 - u_i[k] + I_i[k])\Delta t,
        &\mbox{if $v_i[k] < 30$}, 
        \\
        v_i[k] - (30 - c),&\mbox{otherwise}
    \end{cases}
\end{equation}
and 
\begin{equation}
\label{discrete:2}
        u_i[k+1]
    =    
    \begin{cases}
        u_i[k] + (a(bv_i[k] - u_i[k]))\Delta t,
        &\mbox{if $v_i[k] < 30$}, 
        \\
        u_i[k] + d,&\mbox{otherwise, }
    \end{cases}
\end{equation}
where the input current 
\begin{equation}
    I_i[k] = I_{i, \ctr}[k] + I_{i, \internal}[k]
\end{equation}
consists of the control signal~$I_{i, \ctr}[k]$ to be designed and the input current $I_{i, \internal}[k]$ resulting \add{from firing events at neighbor neurons.}
As in the continuous-time case, we say that a neuron~$i$ fires at time~$k$ if $v_i[k] \geq 30$.

Similar to~\eqref{eq:Iinternal(t)}, we assume the input current $I_{i, \internal}[k]$ of the form 
\begin{equation}
    I_{i, \internal}[k] = \sum_{j\in \mathcal N_i} I_{ij}[k], 
\end{equation}
where $I_{ij}[k]$ represents the current input from neuron~$j$ to neuron~$i$ and is given by 
\begin{equation}\label{eq:I_ij[t]}
    I_{ij}[k] = \begin{cases}
        S(v_j[k])I_{\exitatory},&\mbox{if $j\in \mathcal V_{\exitatory}$}, 
        \\
        S(v_j[k])I_{\inhibitory}, &\mbox{otherwise}
    \end{cases}
\end{equation}
in which the introduced coefficient $S(v_j[t])$ serves the purpose of introducing a soft-threshold and is defined as
\begin{equation}\label{eq:def:Svj}
    S(v_j[k]) = \frac{1}
    {1+\exp(-\sigma(v_j[k]-20))}.
\end{equation}

One of the main reasons for introducing the soft-threshold mechanism is to ensure the differentiability of the objective function. To effectively perform deep learning and deep unfolding, the differentiability of the objective function is an important requirement. 
Conversely, the differentiability of the cost function with respect to the input current $I_{i,\ctr}$ to be designed cannot be guaranteed without soft-thresholding. Therefore, introducing the soft-thresholding mechanism is necessary. In fact, we confirmed through our preliminary experiments that for a system model with an inter-node input current of the form~\eqref{eq:I_ij(t)}, TDU-MPC often fails to yield an effective control input signal.

\subsection{Cost Function}
For reasons similar to soft-thresholding in~\eqref{eq:I_ij[t]} and~\eqref{eq:def:Svj}, the cost function has to be differentiable with the control input to be designed. 
Thus, to design a control input within the first interval~$\mathcal I_1$ in which firing must be promoted within the first module, we adopt an \add{cost} function such that
\begin{equation}
\label{cost_first}
\begin{multlined}[.85\linewidth]
    \sum_{k=l}^{l+T} \biggl(\ \sum_{ i\in\mathcal V_1}(1-\delta_{i}[k]) (30-c - (v_i[k+1]- v_i[k]))  \\ 
     + \sum_{j\in \mathcal V_2}\lvert v_j[k+1] - v_j[k]\rvert\biggr), 
    \end{multlined}
\end{equation}
where 
\begin{equation}
    \delta_{i}[k] = 
    \begin{cases}
        1,&\mbox{if neuron~$i$ fires at time~$k$, }
        \\
        0,&\mbox{otherwise}
    \end{cases}
\end{equation}
is a binary variable encoding the firing event of neurons. 
Minimization of the aforementioned \add{cost} function leads to a smaller first term, which then makes the amount of change in the membrane potential of the neurons in module~$\mathcal V_1$ bigger, thereby stimulating neurons in the module~$\mathcal V_1$ to fire. Conversely, we introduce the second term in~\eqref{cost_first} to suppress firing within the other module~$\mathcal V_2$. 
For the aforementioned reasons, the cost function after the switching time~$T_s$ is set as
\begin{equation}
\label{cost_last}
\begin{multlined}[.85\linewidth]
    \sum_{k=l}^{l+T} \biggl(\ \sum_{i\in \mathcal V_1} \lvert v_i[k+1] - v_i[k]\rvert \\
    + 
    \sum_{ j\in\mathcal V_2}(1-\delta_{j}[k]) (30-c - (v_j[k+1]- v_j[k]))\biggr).
    \end{multlined}
\end{equation}

\begin{figure*}[t]
    \centering
        \begin{minipage}[t]{0.29\hsize}
            \centering
            \includegraphics[keepaspectratio, width=\hsize,trim={1.065in 0.682in 0.913in 0.667in},clip]{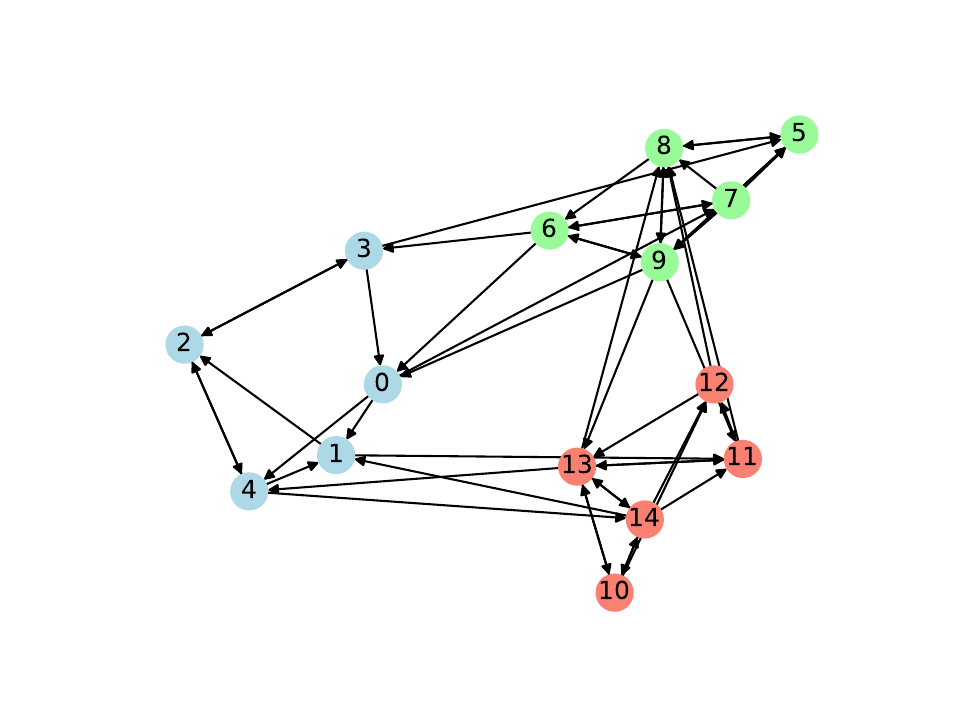}
            \subcaption{Neuronal network~$\mathcal V$. Blue nodes: $\mathcal V_{\ctr}$, green nodes: $\mathcal V_1$, red nodes: $\mathcal V_2$.}
            \label{network:15}
        \end{minipage}
        \hspace{1cm}
        \begin{minipage}[t]{0.29\hsize}
            \centering
            \includegraphics[keepaspectratio, width=\hsize,trim={.228in .0213in .1888in .166in},clip]{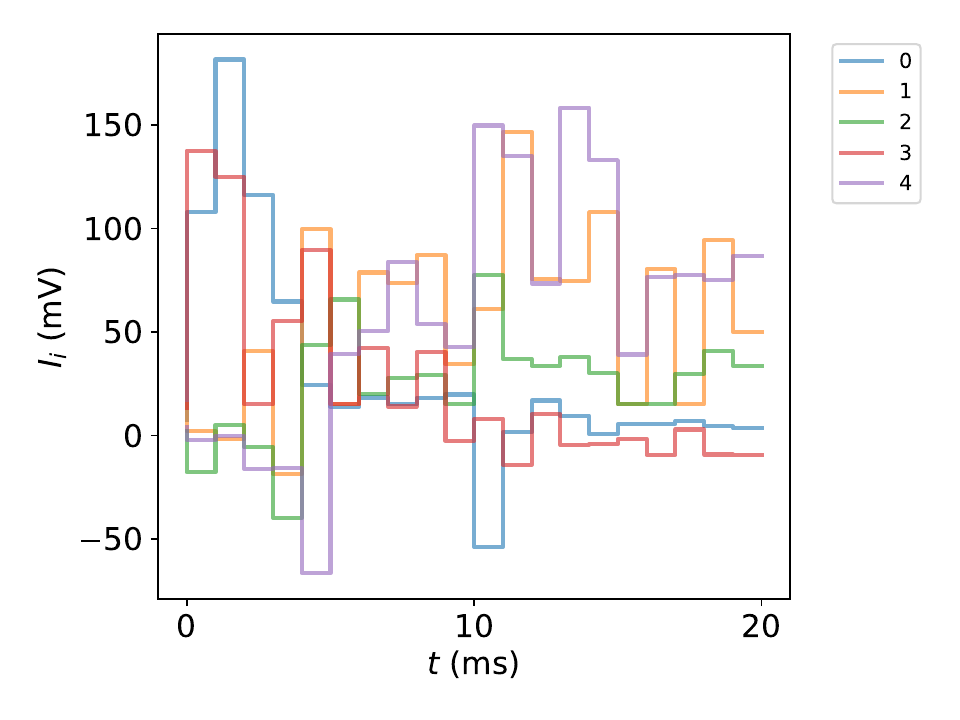}
            \subcaption{Control inputs to neurons in~$\mathcal V_{\ctr}$}
            \label{input1:15}
        \end{minipage}
        \\ \vspace{8mm}
        \begin{minipage}[t]{0.29\hsize}
            \centering
            \includegraphics[keepaspectratio, width=\hsize,trim={.228in .0213in .1888in .166in},clip]{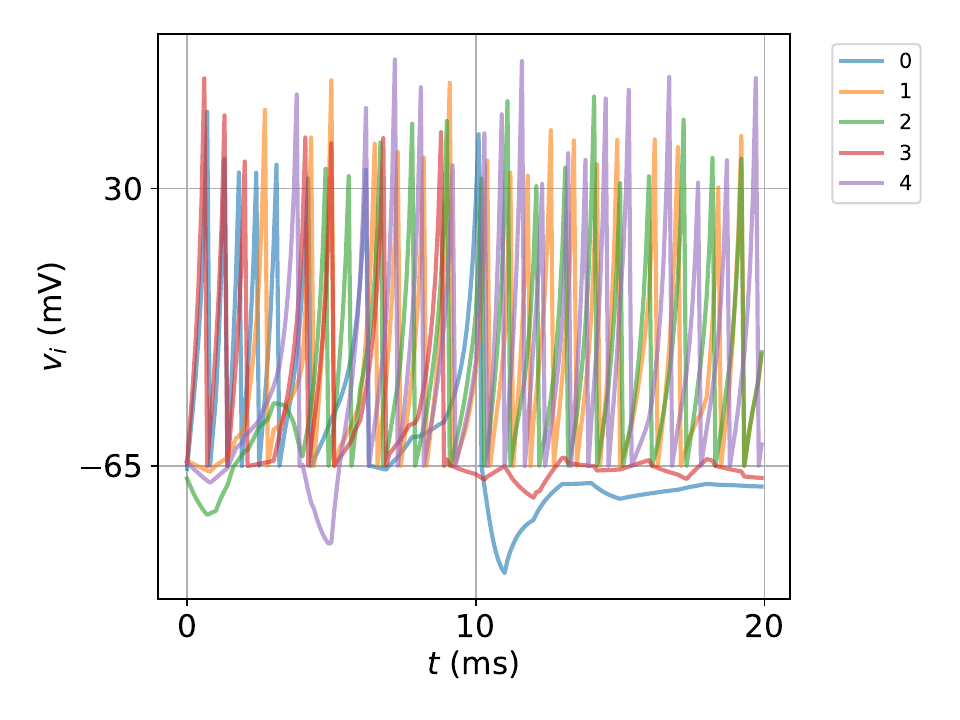}
            \subcaption{Membrane potentials of neurons in~$\mathcal V_{\ctr}$.}
            \label{result0:15}
        \end{minipage}
        \hfill
        \begin{minipage}[t]{0.29\hsize}
            \centering
            \includegraphics[keepaspectratio, width=\hsize,trim={.228in .0213in .1888in .166in},clip]{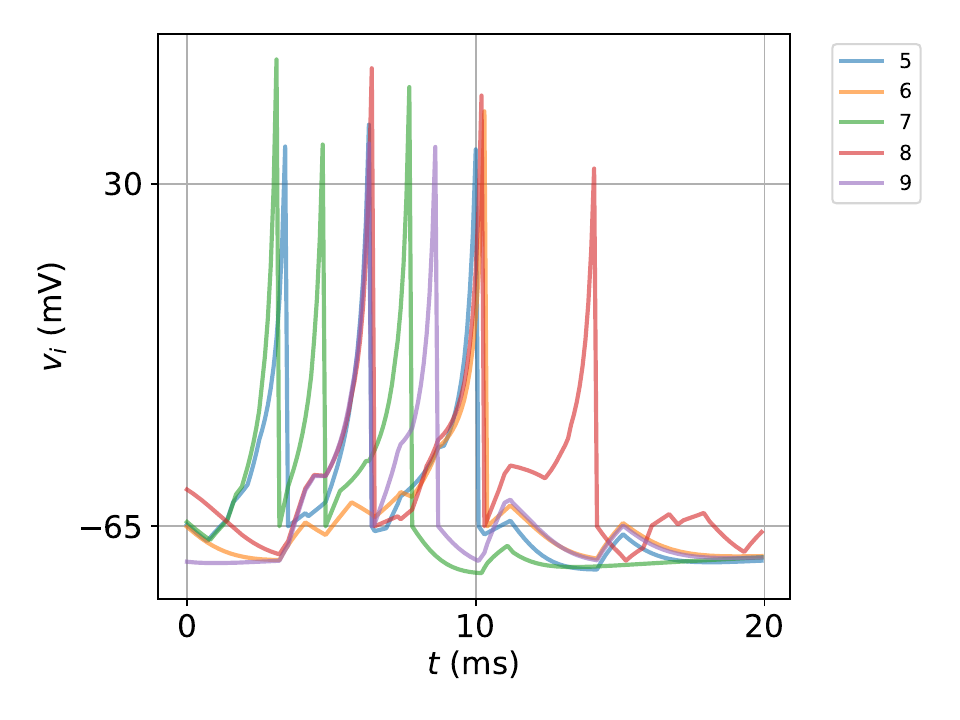}
            \subcaption{Membrane potentials of neurons in~$\mathcal V_1$}
            \label{result1:15}
        \end{minipage}
        \hfill
        \begin{minipage}[t]{0.29\hsize}
            \centering
            \includegraphics[keepaspectratio, width=\hsize,trim={.228in .0213in .1888in .166in},clip]{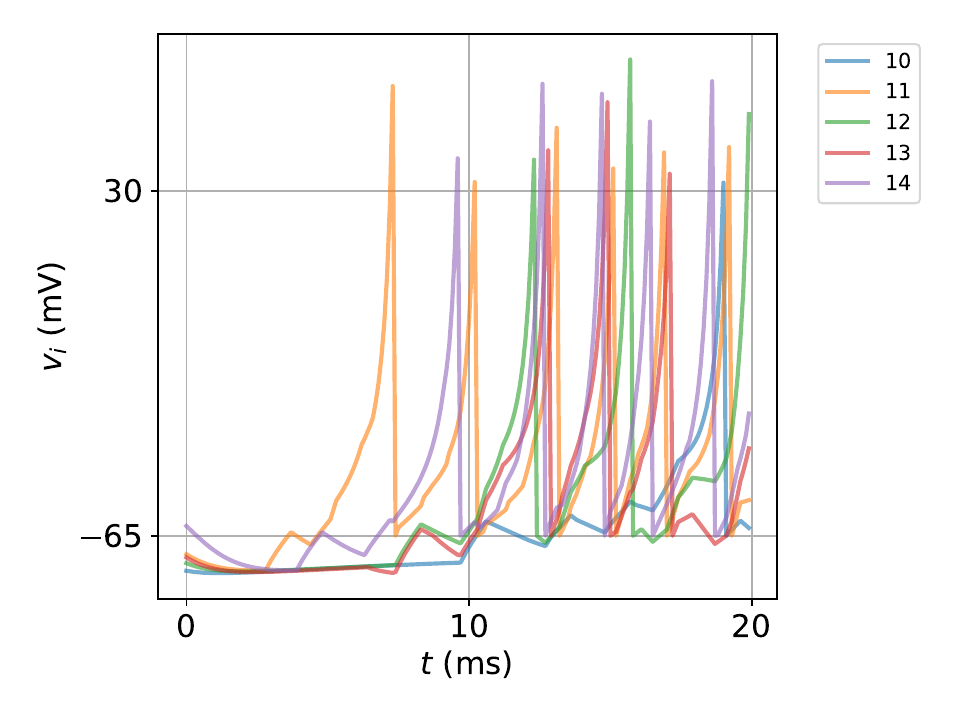}
            \subcaption{Membrane potentials of neurons in~$\mathcal V_2$}
            \label{result2:15}
        \end{minipage}
    \caption{Control of firing pattern in a network with $N=15$ neurons}
    \label{N=15}
    \vspace{5mm}
    \centering
        \begin{minipage}[t]{0.29\hsize}
            \centering
            \includegraphics[keepaspectratio, width=\hsize,trim={1.065in 0.682in 0.913in 0.667in},clip]{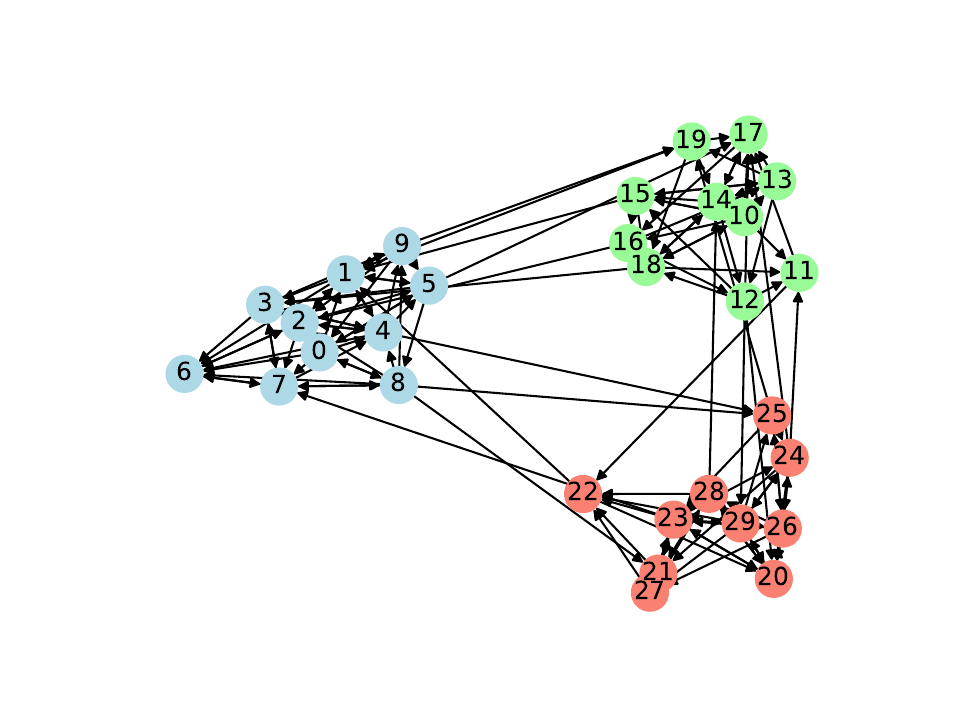}
            \subcaption{Neuronal network~$\mathcal V$. Blue nodes: $\mathcal V_{\ctr}$, green nodes: $\mathcal V_1$, red nodes: $\mathcal V_2$.}
            \label{network:30}
        \end{minipage}
        \hspace{1cm}
        \begin{minipage}[t]{0.29\hsize}
            \centering
            \includegraphics[keepaspectratio, width=\hsize,trim={.228in .0213in .1888in .166in},clip]{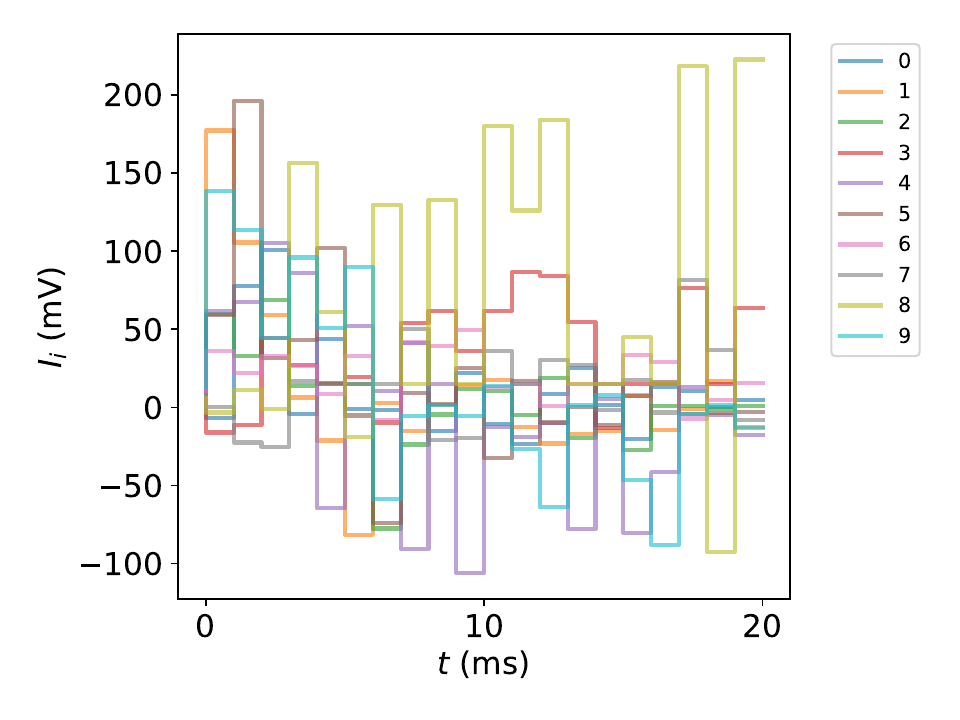}
            \subcaption{Control inputs to neurons in~$\mathcal V_{\ctr}$}
            \label{input1:30}
        \end{minipage}
        \\ \vspace{5mm}
        \begin{minipage}[t]{0.29\hsize}
            \centering
            \includegraphics[keepaspectratio, width=\hsize,trim={.228in .0213in .1888in .166in},clip]{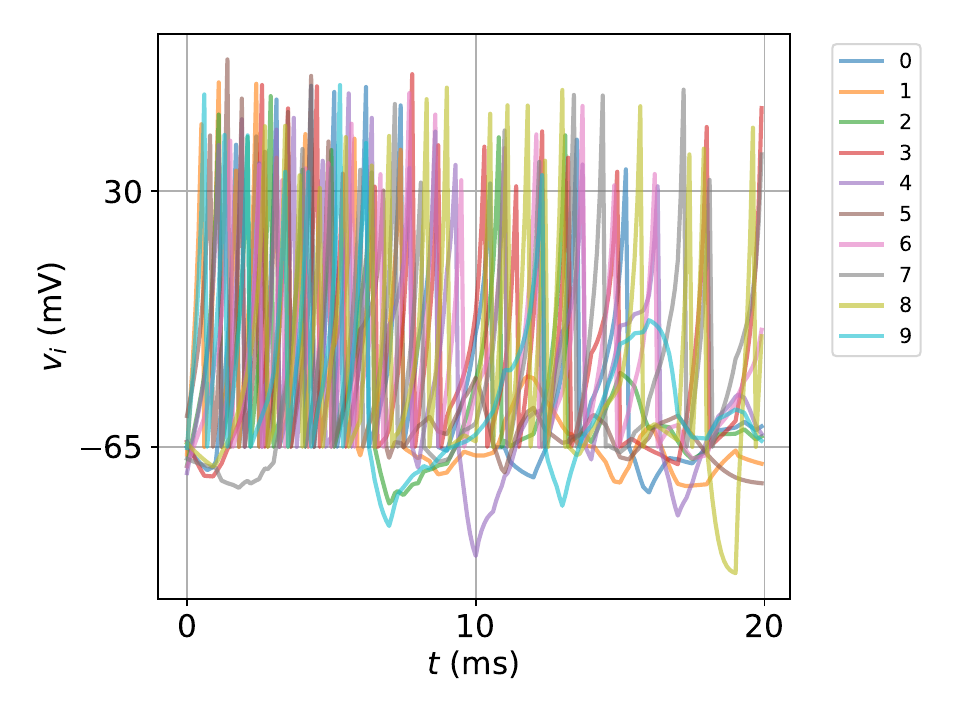}
            \subcaption{Membrane potentials of neurons in~$\mathcal V_{\ctr}$.}
            \label{result0:30}
        \end{minipage}
        \hfill
        \begin{minipage}[t]{0.29\hsize}
            \centering
            \includegraphics[keepaspectratio, width=\hsize,trim={.228in .0213in .1888in .166in},clip]{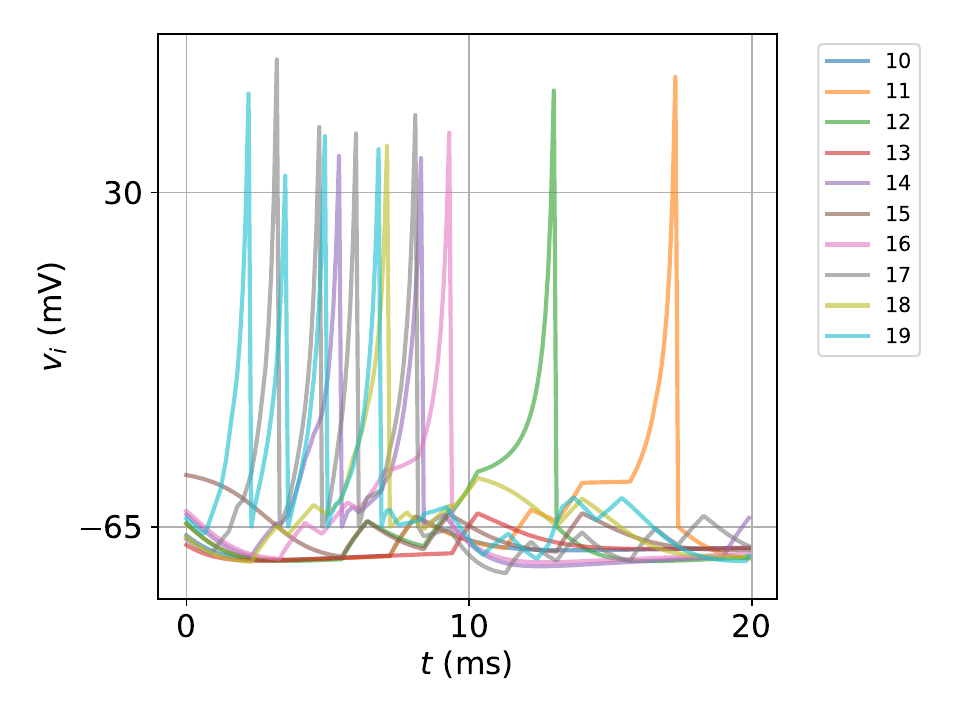}
            \subcaption{Membrane potentials of neurons in~$\mathcal V_1$.}
            \label{result1:30}
        \end{minipage}
        \hfill
        \begin{minipage}[t]{0.29\hsize}
            \centering
            \includegraphics[keepaspectratio, width=\hsize,trim={.228in .0213in .1888in .166in},clip]{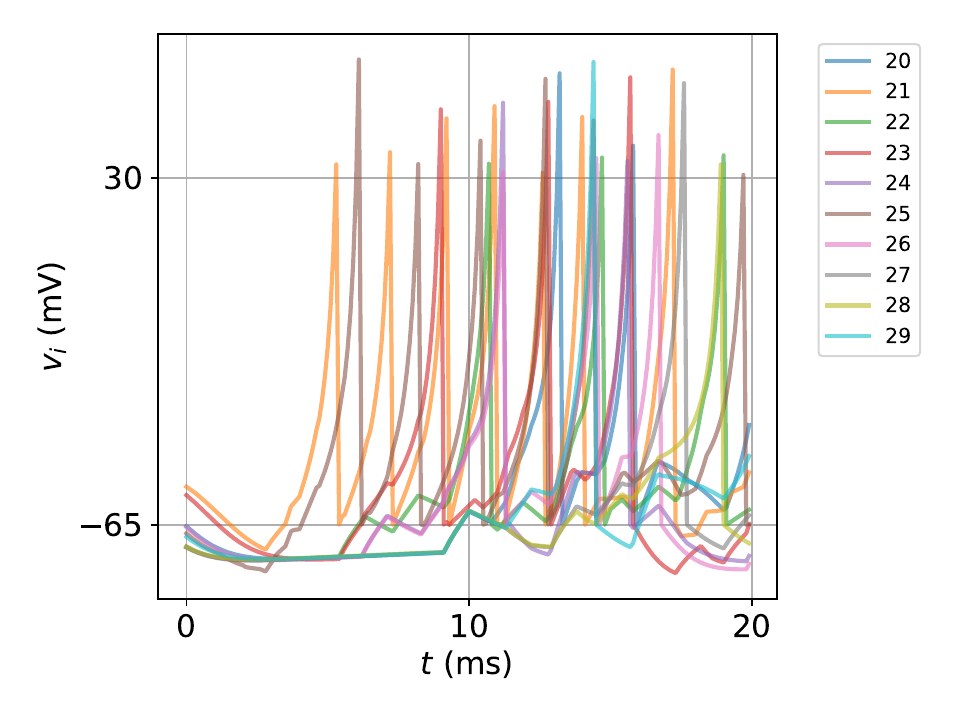}
            \subcaption{Membrane potentials of neurons in~$\mathcal V_2$.}
            \label{result2:30}
        \end{minipage}
    \caption{Control of firing pattern in a network with $N=30$ neurons}
    \label{N=30}
\end{figure*}

\section{Results and Discussions}\label{sec:result}

In this section, we discuss the effectiveness of the proposed approach via numerical simulations. We considered neuronal networks consisting of ``fast spiking'' neurons described in~\cite{izhikevich2003simple}. Therefore, the model parameters in~\eqref{discrete:1} and~\eqref{discrete:2} were set to~$a = 0.1$, $b = 0.2$, $c = -65$, and~$d = 2$. Additionally, we set the amount of current flux incoming from neighboring neurons as $I_{\exitatory} = \SI{15}{\milli\volt}$ and~$I_{\inhibitory} = \SI{-3}{\milli\volt}$. For the neuronal network, we assumed that 80\% of the neurons were of the excitatory type, while 20\% were inhibitory neurons. Further, we assumed that the network consisted of three different modules (or clusters) such as $\mathcal V_1$, $\mathcal V_2$, and~$\mathcal V_\ctr$. As stated in Problem~\ref{prf:}, we aimed to design the control input into the neurons in~$\mathcal V_\ctr$ to control the firing patterns in modules~$\mathcal V_1$ and~$\mathcal V_2$. 

\begin{table*}[tb]
    \centering
    \caption{Numbers of fires in each of the modules}
    \label{firing_time}
        \begin{minipage}[t]{0.45\hsize}
            \centering
            \subcaption{$N=15$}
            \begin{tabular}{|c|c|c|}
            \hline
            \diagbox{Neuron No.}{time (\si{\milli\second})} & 0 $\leq$ time~$<$ 10 & 10 $\leq$ time~$\leq$ 20 \\ \hline \hline
            5-9        & \textbf{8}                             & 4                               \\ \hline
            10-14      & 2                             & \textbf{16}                                \\ \hline
            \end{tabular}
            \label{firing_time:15}
        \end{minipage}
        \hspace{.75cm}
        \begin{minipage}[t]{0.45\hsize}
            \centering
            \subcaption{$N=30$}
            \begin{tabular}{|c|c|c|}
            \hline
            \diagbox{Neuron No.}{time (\si{\milli\second})} & 0 $\leq$ time~$<$ 10 & 10 $\leq$ time~$\leq$ 20 \\ \hline \hline
            10-19        & \textbf{12}                             & 2                               \\ \hline
            20-29      & 6                             & \textbf{23}                                \\ \hline
            \end{tabular}
            \label{firing_time:30}
        \end{minipage}
\end{table*}

The (hyper-)parameters of the proposed method were chosen as follows. First, Adam was chosen as the optimizer, an algorithm used for updating the control input, based on existing literature~\cite{bae2019does}. Second, similar to ~\cite{kishida2021}, incremental learning was adopted during the process of control input learning. Third, for the Python implementation of TDU-MPC, we used libraries such as PyTorch~\cite{paszke2019pytorch} and NumPy~\cite{van2011numpy}. Last, the time step for discretization was set to~$\Delta t = \SI{1}{\milli\second}$\masaki{相澤君，単位を明示してください．}, while the prediction horizon within TDU-MPC was set as $T = 10$ \add{step}\masaki{相澤君，単位を明示してください．}. Furthermore, for the soft-thresholding function~\eqref{eq:def:Svj}, we chose parameter~$\sigma = 0.38$.

As aforementioned, we first generated a network with $N=15$ neurons and performed a control experiment via numerical simulations. To construct a network with modules, we used the stochastic block model~\cite{lee2019review} with Modules~$\mathcal V_\ctr = \{0, 1, 2, 3, 4\}$, $\mathcal V_1 = \{5, 6, 7, 8, 9\}$, and~$\mathcal V_2 = \{10, 11, 12, 13, 14\}$. \add{Out of the $15$ neurons, neurons~$6$, $9$, and $13$\masaki{相澤君，inhibitory 
 neuronのindexをここに書いて下さい．下も同様です．} were inhibitory, while the others were excitatory.} The edge probability within the module was set to~$1/2$, while the one between modules was set to~$1/8$. Utilizing these parameters, we obtained the network shown in Fig.~\ref{network:15}. \waka{接続数について説明があると良い}\masaki{done}
\add{Within the resulting network, although the number of connections between the modules are fewer than the ones within the modules, the inter-module connections are still bidirectional, which makes the control problem nontrivial.}

For initializing the membrane potentials~$v_i$ and membrane recovery variables~$u_i$, we first set the potentials to~$c$ and the recovery variables to~$0$. Thereafter, a random current was applied to all neurons for \SI{10}{\milli\second}. After initialization, we performed neuronal network control using the method described in Section~\ref{sec:proposed}. The switching and simulation times were set to~$T_s = \SI{10}{\milli\second}$ and~$T_e = \SI{20}{\milli\second}$, respectively. Our objective was to design control inputs $I_{0,\ctr}$, \dots, $I_{4,\ctr}$ such that the firing pattern of~$\mathcal V_1$ was more active than that of~$\mathcal V_2$ in the time interval $\mathcal I_1 = [\SI{0}{\milli\second}, \SI{10}{\milli\second}]$, and that of~$\mathcal V_2$ was more active than $\mathcal V_1$ in the time interval $\mathcal I_2 = [\SI{10}{\milli\second}, \SI{20}{\milli\second}]$.

The results of the control experiment are presented in Fig.~\ref{N=15}.
In Fig.~\ref{input1:15}, the control input signals to neurons in~$\mathcal V_\ctr$ obtained from our method have been illustrated. From the membrane potentials of neurons in~$\mathcal V_1$ and~$\mathcal V_2$ shown in Figs.~\ref{result1:15} and~\ref{result2:15}, respectively, we confirmed that module~$\mathcal V_1$ fired frequently from the initial time to~$t=\SI{10}{\milli\second}$ as compared with $\mathcal V_2$. Additionally, module~$\mathcal V_2$ fired frequently from $t=\SI{10}{\milli\second}$ to~$t=\SI{20}{\milli\second}$ when compared with $\mathcal V_1$, as desired. We observed that the average computation time for a control input for the duration of \SI{1}{\milli\second} was \SI{106}{\second}. Thus, improvements to the current implementation for a shorter computational time are of practical importance.

Further, the simulation results with $N=30$ neurons were observed. The modules were set as $\mathcal V_\ctr = \{0, 1, 2, \dotsc, 9\}$, $\mathcal V_1 = \{10, 11, 12, \dotsc, 19 \}$, and~$\mathcal V_2 = \{20, 21, 22, \dotsc, 29\}$. \add{Out of the $30$ neurons, neurons~$3$, $4$, $6$, $20$, $26$, and $27$ were inhibitory, while the others were excitatory.} The edge probability within the network was set to~$1/2$, while the one between modules was set to~$1/25$. With these parameters, we constructed a neuronal network with 30 neurons, as shown in Fig.~\ref{network:30}. 
Similar to our numerical experiment for the network with $N=15$ nodes, we designed the control input~$I_{i, \ctr}$ ($i\in \mathcal V_\ctr$) for the new network. The obtained time series of the control input has been shown in Fig.~\ref{input1:30}. Further, the membrane potentials of~$\mathcal V_1$ and~$\mathcal V_2$ under the control input are shown in Figs.~\ref{result2:30} and~\ref{result2:30}, respectively. Modules~$\mathcal V_1$ and~$\mathcal V_2$ were confirmed to fire in the desired manner, and the average computational time for one step of the control input was \SI{353}{\second}.

The results presented in Table~\ref{firing_time} provide a summary of our numerical experiment for the two networks. Within the table, the number of fires in each of the modules for time intervals~$\mathcal I_1$ and~$\mathcal I_2$. The number of fires within the targeted module (the one required to be more active than the other) was confirmed to be no less than double the number of the one within the other module, thereby illustrating the effectiveness of the proposed method. 

\waka{これ以降、図の説明だけで考察がない。例えば上記のコメントしたようにVcontrolからV1、V2への接続は少ない、同時にVcontrolはV1、V2から影響を受ける。なぜこの状況で制御できているのか？
V1を頻繁に発火させる時とV2を頻繁に発火させる時では入力はどう違うのか、なぜそうなるのか。単にMPCがそう決めたからではなく、その意味をちゃんと解釈すること。
イントロで言っていることに対してニューロン数が少ない。100、1000，10000と増えた時にどうなるか。}\masaki{相澤君：ところで以下の記述は全てのニューロンがexcitatoryであるような感じになっています．$V_\ctr$のニューロンが全てexcitatoryだったりしますか？あと図に関しては各ニューロンがexcitatoryかinhibitoryか図から読み取れると理想的だよね．}
\add{In both simulation results, initial time interval~$\mathcal I_1$ corresponds to the period where module~$\mathcal V_1$ should display higher firing activity. During this interval, the applied control input primarily activated the excitatory neurons in~$V_\ctr$ having edges outgoing to~$\mathcal V_1$. For instance, when considering $N=15$ as an example, we can observe from Fig.~\ref{input1:15} that neurons~$0$ and~$3$ were predominantly stimulated at the beginning of the time interval. As the switching time~$T_s$ approaches, the control input gradually diminished, effectively preventing firing within~$\mathcal V_1$ in the subsequent time interval~$\mathcal I_2$ when $\mathcal V_2$ was expected to exhibit high firing activity. Similarly, in the second time interval~$\mathcal I_2$, we observed a comparable pattern to the aforementioned case. The excitatory neurons in~$V_\ctr$ having edges outgoing to~$\mathcal V_2$ were primarily stimulated. For instance, considering the case of~$N=15$ again, neurons~$1$ and~$4$ demonstrated increased activation during this time interval. These observed stimulation patterns align well with our intuitive understanding and expectations.}


\section{Conclusions}\label{sec:conclusion}

This study addressed the challenges associated with neuronal network control. We proposed a novel method for designing control inputs to manipulate the firing patterns of modules within a network. Based on TDU-MPC, our approach exploits deep unfolding techniques commonly used in wireless signal processing. The effectiveness of our proposed method was demonstrated via extensive numerical simulations.
\aizawa{In networks with 15 and 30 neurons, the control was achieved to switch the firing frequencies of two modules without directly applying control inputs.}

Future research directions include evaluating the effectiveness of the proposed method via numerical experiments on other networks consisting of neurons other than the fast-spiking type. Additionally, robustness evaluation of the proposed method with respect to the modeling errors of the neurons can be of practical relevance\add{, as the current research is conducted on the assumption that the dynamics of  neurons are available and, therefore, errors in the model are assumed to be not significant}. \add{Another important research direction is the investigation of a method to design control inputs without the need to observe all membrane potentials.}\edi{研究の限界を含めても良い}\masaki{done}
\balance




\begin{thebibliography}{1}

\bibitem{hutchison2013dynamic}
R.M.~Hutchison, T.~Womelsdorf, E.A.~Allen, P.A.~Bandettini, V.D.~Calhoun, M.~Corbetta, S.D.~Penna, et.al,
``Dynamic functional connectivity: promise, issues, and interpretations'',
\emph{Neuroimage}, 
vol.~80, pp.~360--378, 2013.

\bibitem{tang2018colloquium}
E.~Tang and D.S.~Bassett,
``Colloquium: Control of dynamics in brain networks'',
\emph{Reviews of Modern Physics}, 
vol.~90, 2018.

\bibitem{lee2019current}
D.J.~Lee, C.S.~Lozano, R.F.~Dallapiazza and A.M.~Lozano,
``Current and future directions of deep brain stimulation for neurological and psychiatric disorders: JNSPG 75th Anniversary Invited Review Article'',
\emph{Journal of Neurosurgery}, 
vol.~131, pp.~333--342, 2019.

\bibitem{wander2014brain}
J.D.~Wander and R.P.N.~Rao,
``Brain--computer interfaces: a powerful tool for scientific inquiry'',
\emph{Current Opinion in Neurobiology}, 
vol.~25, pp.~70--75, 2014.

\bibitem{iolov2014stochastic}
A.~Iolov, S.~Ditlevsen and A.~Longtin, ``Stochastic optimal control of single neuron spike trains,'' \emph{Journal of Neuronal Engineering}, vol.~11, 2014.

\bibitem{izhikevich2003simple}
E.M.~Izhikevich, ``Simple model of spiking neurons'' \emph{IEEE Transactions on Neuronal Networks}, vol.~14, pp.~1569--1572, 2003.

\bibitem{motter2015networkcontrology}
A.E.~Motter,
``Networkcontrology'',
\emph{Chaos: An Interdisciplinary Journal of Nonlinear Science}, 
vol.~25, 2015.

\bibitem{kishida2021}
M.~Kishida and M.~Ogura, ``Temporal deep unfolding for constrained nonlinear stochastic optimal controls,'' \emph{{IET Control Theory \& Applications}}, vol.~16, pp.~139--150, 2022.

\bibitem{hershey2014deep}
J.R.~Hershey, J.L.~Roux, and F.~Weninger, ``Deep unfolding: Model-based
  inspiration of novel deep architectures,'' {arXiv:1409.2574}, 2014.
  
\bibitem{Wadayama}
D.~Ito, S.~Takabe, and T.~Wadayama, ``Trainable ISTA for sparse signal recovery,'' \emph{IEEE Transactions on Signal Processing}, vol.~67, pp.~3113--3125, 2019.

\bibitem{Monga}
V.~Monga, Y.~Li and Y.C.~Eldar, ``Algorithm unrolling: Interpretable efficient deep learning for signal and image processing,'' \emph{IEEE Signal Processing Magazine}, vol.~38, pp.~18--44, 2021.

\bibitem{Melodia}
A.~Jagannath, J.~Jagannath and T.~Melodia, ``Redefining wireless communication for 6G: Signal processing meets deep learning with deep unfolding,'' \emph{IEEE Transactions on Artificial Intelligence}, vol.~2, pp.~528--536, 2021.

\bibitem{Ogura}
M.~Ogura, K.~Kobayashi, and K.~Sugimoto, ``Static output feedback synthesis of time-delay linear systems via deep unfolding,'' In \emph{17th IFAC Workshop on Time Delay Systems}, pp.~214--215, 2022.

\bibitem{carrasco2010passivity}
J.~Carrasco, A.~Ba{\~n}os and A van der Schaft,
``A passivity-based approach to reset control systems stability'',
\emph{Systems \& Control Letters}, 
vol.~59, pp.~18--24, 2010.

\bibitem{garcia1989model}
C.E.~Garcia, D.M.~Prett and M.~Morari,
``Model predictive control: Theory and practice—A survey'',
\emph{Automatica}, 
vol.~25, pp.~335--348, 1989.

\bibitem{bae2019does}
K.~Bae, H.~Ryu, and H.~Shin, ``Does adam optimizer keep close to the optimal point?'' \emph{arXiv preprint arXiv:1911.00289}, 2019.

\bibitem{paszke2019pytorch}
A.~Paszke, S.~Gross, F.~Massa, A.~Lerer, J.~Bradbury et al., ``Pytorch: An imperative style, high-performance deep learning library'' \emph{Advances in Neuronal Information Processing Systems}, vol.~32, 2019.

\bibitem{van2011numpy}
S.~Van Der Walt, S.C.~Colbert and G.~Varoquaux, 
``The NumPy array: a structure for efficient numerical computation'' 
\emph{Computing in Science \& Engineering}, 
vol.~13, pp.~22--30, 2011.

\bibitem{lee2019review}
C.~Lee and D.J.~Wilkinson,
``A review of stochastic block models and extensions for graph clustering'',
\emph{Applied Network Science}, 
vol.~4, pp.~1--50, 2019.


  
\end{thebibliography}

\end{document}